\documentclass[12pt,reqno]{article}
\usepackage{amsmath}

\allowdisplaybreaks
\numberwithin{equation}{section}

\newcommand{\calL}{\mathcal{L}}
\renewcommand{\d}{\mathrm{d}}
\newcommand{\I}{\mathrm{i}}
\newcommand{\M}{\mathcal{M}}
\newcommand{\e}{\mathrm{e}}
\newcommand{\ep}{\epsilon}
\newcommand{\p}{\partial}

\newcommand{\w}{\wedge}
\newcommand{\half}{\tfrac{1}{2}}
\newcommand{\frc}[2]{\frac{\raisebox{-2pt}{$#1$}}{#2}}
\newcommand{\tbs}[1]{\overset{\leftrightarrow}{#1}}

\DeclareMathOperator{\im}{Im}
\DeclareMathOperator{\re}{Re}

\DeclareSymbolFont{AMSa}{U}{msa}{m}{n}
\DeclareSymbolFont{AMSb}{U}{msb}{m}{n}
\DeclareMathSymbol{\fieldR}{\mathalpha}{AMSb}{"52}

\begin{document} 

\begin{flushright} \small
 ITP--UU--02/43 \\ SPIN--02/25 \\ hep-th/0208145
\end{flushright}
\bigskip

\begin{center}
 {\large\bfseries Instantons in the Double-Tensor Multiplet} \\[5mm]
 Ulrich Theis and Stefan Vandoren \\[2mm]
 {\small\slshape
 Spinoza Instituut, Universiteit Utrecht \\
 Postbus 80.195, 3508 TD Utrecht, The Netherlands \\[2pt]
 U.Theis, S.Vandoren@phys.uu.nl}
\end{center}
\vspace{5mm}

\hrule\bigskip

\centerline{\bfseries Abstract} \medskip

The double-tensor multiplet naturally appears in type IIB superstring
compactifications on Calabi-Yau threefolds, and is dual to the universal
hypermultiplet. We revisit the calculation of instanton corrections to
the low-energy effective action, in the supergravity approximation. We
derive a Bogomol'nyi bound for the double-tensor multiplet and find new
instanton solutions saturating the bound. They are characterized by the
topological charges and the asymptotic values of the scalar fields in
the double-tensor multiplet.
\bigskip

\hrule\bigskip

\section{Introduction} 

Instanton effects in string and M-theory are still relatively poorly
understood. This is due to the lack of a conventional instanton calculus
as we know it from (supersymmetric) field theory. A well-known open
problem is to determine the instanton corrections to the hypermultiplet
moduli space of type II superstrings or M-theory compactified on a
Calabi-Yau (CY) threefold down to four or five dimensions. Supersymmetry
requires the hypermultiplet moduli space ${\cal M}_H$ to be
quaternion-K\"ahler \cite{BW}. The four- (or five-) dimensional dilaton
lives in a multiplet which can be dualized into the universal
hypermultiplet. Hence, ${\cal M}_H$ receives quantum corrections, and
the instantons correspond to Euclidean $p$-branes wrapping $p+1$ cycles
of the CY \cite{BBS}.

The simplest setup for studying this problem, is to consider
CY-compactifications of M-theory/type IIA superstrings with Hodge number
$h_{2,1}=0$, or, for type IIB, $h_{1,1}=0$~\footnote{By type IIB with
$h_{1,1}=0$ we mean the mirror version of IIA with $h_{1,2}=0$. As
explained in \cite{AG}, this model has to be understood in terms of a
Landau-Ginzburg description instead of a geometric compactification,
since all CY manifolds are K\"ahler and have $h_{1,1}>0$.}, since this
yields a low-energy effective action of $N=2$ supergravity coupled to a
single hypermultiplet, such that the moduli space ${\cal M}_H$ has
dimension four. From a type IIB perspective, this hypermultiplet arises
from dualizing the \emph{double-tensor} multiplet, whose bosonic
components descend from the $NS$-$NS$ and $R$-$R$ two-forms and scalars
in ten dimensions. This suggests that instanton calculations should be
done on the double-tensor multiplet side. In the next section, we shall
make another argument, which also applies to type IIA and M-theory, why
the double-tensor multiplet is more appropriate for our purposes.

Yet, even in the case of a single hypermultiplet, it is difficult to
compute instanton effects directly in string theory, without explicit
knowledge of the instanton measure and the details of the wrapped branes
along the CY cycles. Therefore, we will study this problem in a pure
supergravity context, in which semi-classical instanton calculations can
be done in the more conventional and ``field-theoretic'' way, following
a similar strategy as in \cite{BB,GS1}, or as in \cite{R} for matter
coupled to $N=1$ supergravity. Although being an approximation of the
exact result, the hope is that the leading supergravity corrections,
combined with the constraints from quaternion-K\"ahler geometry, and
together with some knowledge from string theory on the isometries and
singularity structure of ${\cal M}_H$, should fix the answer uniquely.
Such a program has worked succesfully in the context of supersymmetric
field theories in three dimensions with eight supercharges, where the
hypermultiplet moduli space is hyperk\"ahler \cite{SW,DKMTV}. See
\cite{OV,K} for related issues.

In this paper, we carry out the first steps of the supergravity
instanton calculation. In section 2, we explain how the Euclidean theory
is best understood in terms of the double-tensor multiplet, since then
the action is manifestly positive definite, a requirement needed for a
semiclassical approximation. In section 3, we derive a Bogomol'nyi bound
and show that the instanton action is purely topological and given by a
surface term. We then solve the BPS equation explicitly and compute the
instanton action for the solutions. A similar approach was followed in
\cite{BB} and \cite{GS1}. Compared to these papers, we propose a
different Euclidean version of the universal hypermultiplet Lagrangian,
so our results, where comparable, are somehow different. Moreover, we
have found new instanton solutions, which will play an important role in
understanding the quantum corrected hypermultiplet moduli space, as
explained in the discussion at the end of the paper.

\section{The double-tensor multiplet} 

As mentioned in the introduction, we are interested in the case of a
single hypermultiplet coupled to $N=2$ supergravity. Classically, the
four scalars of the universal hypermultiplet parametrize the
homogeneous quaternion-K\"ahler manifold \cite{CFG,FS}
 \begin{equation} \label{UHM-QK}
  \M_H = \frc{\mathrm{SU}(1,2)}{\mathrm{U}(2)}\ .
 \end{equation}
In a basis of real fields $\{\phi,\chi,\varphi,\sigma\}$, the bosonic
Lagrangian takes the form\footnote{Throughout this paper, we use
form notation. Lagrangians are written as volume forms, and we use the
notation $|\omega_p|^2\equiv *\omega_p\w\omega_p$.}
 \begin{equation} \label{UHM-action}
  \calL_\mathrm{UH} = - \d^D x\, \sqrt{g\,} R + \half |\d\phi|^2 + \half\,
  \e^{-\phi} \big( |\d\chi|^2 + |\d\varphi|^2 \big) + \half\,
  \e^{-2\phi}\, |\d\sigma + \chi \d\varphi|^2\ ,
 \end{equation}
with $D=4$ or $5$, depending on whether one is interested in type II or
M-theory compactifications. The Lagrangian has a global SU(1,2)
isometry group.

For our purposes, it will be convenient to discuss the dual version of
$\calL_\mathrm{UH}$ in terms of a double-tensor multiplet. Consider  the
first-order Lagrangian
 \begin{equation} \label{DTM-action}
  \calL_\mathrm{DT} = - \d^Dx\, \sqrt{g\,} R + \half |\d\phi|^2 + \half\,
  \e^{-\phi} |\d\chi|^2 + \half M_{ab} *\! H^a \w H^b - \lambda_a\,
  \d H^a\ ,
 \end{equation}
where the $H^a$ are a doublet of $(D-1)$ forms, the $\lambda_a$ are two
scalars, and
 \begin{equation}
  M = \e^{\phi} \begin{pmatrix} 1 & - \chi \\[2pt] - \chi & \e^{\phi}
  + \chi^2 \end{pmatrix}\ .
 \end{equation}
The two scalars $\phi$ and $\chi$ parametrize the coset SL$(2,\fieldR)/
\mathrm{O}(2)$; in terms of the complex combination 
 \begin{equation} \label{tau}
  \tau \equiv \chi + 2\I\, \e^{\phi/2}
 \end{equation}
the scalar part of $\calL_\mathrm{DT}$ can be written as $2|d\tau/\im
\tau|^2$. The tensor terms, however, break the global SL$(2,\fieldR)$
symmetry, leaving only shift symmetries of $\phi$ and $\chi$. The shift
in $\chi$ acts as
 \begin{equation}\label{shift-chi}
  \tau \rightarrow \tau + b\ ,\quad \begin{pmatrix} H^1 \\[2pt] H^2
  \end{pmatrix} \rightarrow \begin{pmatrix} 1 & b \\[2pt] 0 & 1
  \end{pmatrix} \begin{pmatrix} H^1 \\[2pt] H^2 \end{pmatrix}\ ,
  \quad \begin{pmatrix} \lambda^1 \\[2pt] \lambda^2 \end{pmatrix}
  \rightarrow \begin{pmatrix} 1 & 0 \\[2pt] -b & 1 \end{pmatrix}
  \begin{pmatrix} \lambda^1 \\[2pt] \lambda^2 \end{pmatrix}\ ,
 \end{equation}
whereas the shift in $\phi$ acts as
 \begin{equation}\label{shift-phi}
  \tau \rightarrow \e^\kappa \tau\ ,\quad \begin{pmatrix} H^1 \\[2pt]
  H^2 \end{pmatrix} \rightarrow \begin{pmatrix} \e^{-\kappa} H^1
  \\[2pt] \e^{-2\kappa} H^2 \end{pmatrix}\ ,\quad \begin{pmatrix}
  \lambda^1 \\[2pt] \lambda^2 \end{pmatrix} \rightarrow \begin{pmatrix}
  \e^\kappa \lambda^1 \\[2pt] \e^{2\kappa} \lambda^2 \end{pmatrix}\ .
 \end{equation}
Note that the latter acts like an SL$(2,\fieldR)$ transformation on
$\tau$, but not on the $H^a$. The full type IIB theory compactified to
four dimensions (classically) has SL$(2,\fieldR)$ symmetry, due to the
presence of additional tensor multiplets which transform nontrivially
\cite{BGHL}. Setting the scalars in these multiplets to nonvanishing
constants results in a breakdown of the symmetry and leaves only the
above transformations as residual invariances.

The equations of motion for the Lagrange multipliers $\lambda_a$ imply
that the $H^a$ are closed. Writing $H^a=\d B^a$, one obtains the
double-tensor multiplet. Integrating out the tensors instead gives the
duality relation
 \begin{equation} \label{dual-rel}
  \d\lambda_a = - M_{ab} *\! H^b\ .
 \end{equation}
Substituting this back yields the action for the universal
hypermultiplet \eqref{UHM-action}, upon identifying
 \begin{equation} \label{mult1}
  \lambda_1 = \varphi\ ,\quad \lambda_2 = \sigma\ .
 \end{equation}
The dual formulation in terms of the double-tensor multiplet is not
unique\footnote{This fact was already observed in \cite{DWRV}. The
reason is that one can choose inequivalent commuting $U(1)\times U(1)$
factors in the isometry group SU(1,2) with respect to which we dualize
the universal hypermultiplet into the double-tensor multiplet. The
choices above correspond to shifts in $\varphi$ and $\sigma$, and a
shift in $\sigma$ together with the transformation $\chi\rightarrow
\chi+\epsilon$, $\sigma\rightarrow\sigma-\epsilon\varphi$.}. We can
start with \eqref{DTM-action}, but write everywhere $\varphi$ instead
of $\chi$. Dualizing the tensors and identifying
 \begin{equation} \label{mult2}
  \lambda_1 = - \chi\ ,\quad \lambda_2 = \sigma + \varphi\, \chi
 \end{equation}
yields the same hypermultiplet action, as one can easily check.

In addition, the dualization procedure yields a boundary term which has
to be added to the hypermultiplet action,
 \begin{equation} \label{surf-term}
  \calL_\mathrm{bnd} = (-)^D\, \d \big[ \lambda_a\, (M^{-1})^{ab} *\d
  \lambda_b\big]\ ,
 \end{equation}
where we used that, when acting on a $p$-form in Minkowski space, $**=
-(-)^{(D-1)p}$. The different choices corresponding to \eqref{mult1} and
\eqref{mult2} would now give different boundary terms. However, due to
the isometries of the scalar manifold, they are related to each other by
a field redefinition of the multipliers, $\tilde{\sigma}=\sigma+\varphi
\chi$, $\tilde{\varphi}=-\chi$, $\tilde{\chi}=\varphi$. Substituting
\eqref{mult2} into \eqref{surf-term}, we get
 \begin{equation}
  \calL_\mathrm{bnd} = (-)^D\, \d \big[ \e^{-\phi} \chi *\! \d\chi +
  \e^{-2\phi} \sigma *\! (\d\sigma + \chi \d\varphi) \big]\ .
 \end{equation}
The total action for the universal hypermultiplet is then
 \begin{equation}
  \calL = \calL_\mathrm{UH} + \calL_\mathrm{bnd}\ .
 \end{equation}

The fermions have been suppressed here. For hypermultiplets, the
supersymmetry transformation rules and the fermion-terms in the
Lagrangian are known in general. For the double-tensor multiplet
Lagrangian \eqref{DTM-action}, the fermion-terms and susy rules can be
determined by dualization. However, the most general self-interacting
supersymmetric double-tensor multiplet Lagrangian has not been worked
out. For a discussion on this in the context of rigid $N=2$
supersymmetry, we refer to \cite{B}.

\subsection*{Euclidean formulation} 

To find instanton solutions, we need the Euclidean formulation of the
universal hypermultiplet, or, equivalently, the Euclidean double-tensor
multiplet Lagrangian. For the latter, apart from the usual complications
with the Euclidean Einstein-Hilbert term, the Wick rotation acts in the
standard way on the scalars and tensors. While the double-tensor
multiplet Lagrangian formally stays the same,
 \begin{equation} \label{E-DTM}
  \calL_\mathrm{DT}^E = \d^Dx\, \sqrt{g\,} R + \half |\d\phi|^2 + \half\,
  \e^{-\phi} |\d\chi|^2 + \half M_{ab} *\! H^a \w H^b\ ,
 \end{equation}
the dual Euclidean universal hypermultiplet Lagrangian has two sign flips
in the kinetic terms, due to the fact that we now have $**=(-)^{(D-1)p}$
when acting on a $p$-form in Euclidean space. The dualization procedure
yields
 \begin{equation} \label{EUHM-action}
  \calL_\mathrm{UH}^E = \d^D x\, \sqrt{g\,} R + \half |\d\phi|^2 + \half\,
  \e^{-\phi} \big( |\d\chi|^2 - |\d\varphi|^2 \big) - \half\, \e^{-2
  \phi}\, |\d\sigma + \chi \d\varphi|^2\ ,
 \end{equation}
together with the boundary term
 \begin{equation} \label{E-BT}
  \calL_\mathrm{bnd}^E = - (-)^D\, \d \big[ \e^{-\phi} \chi *\! \d\chi +
  \e^{-2\phi} \sigma *\! (\d\sigma + \chi \d\varphi) \big]\ .
 \end{equation}
By setting $\varphi=\chi=0$, this boundary term is the same as for the
$D$-instanton of type IIB in ten dimensions, obtained by dualizing the
nine-form field strength into the $R$-$R$ scalar $\sigma$ \cite{GGP,GG}.
In four dimensions, we generate more terms due to the fact that we
dualize two tensors. The sign flips of the kinetic terms of the two dual
fields $\lambda_a$ are compatible with the prescription of Wick rotating
pseudoscalars $\lambda_a\rightarrow\I\lambda_a$ \cite{vNW}. This is
consistent with the duality relation \eqref{dual-rel}.

A Euclidean version of the universal hypermultiplet action was also
proposed in \cite{GS1}. Both their bulk Lagrangian and boundary term
differ from ours. This has important consequences since the instanton
action defines the weight in the path integral, and hence correlation
functions and eventually the quantum-corrected hypermultiplet moduli
space will be different.

Due to the sign changes in \eqref{EUHM-action}, the geometry of the
scalar manifold is no longer SU(1,2)/U(2). Instead, it is given by the
coset space
 \begin{equation} \label{SL3}
  \M_H^E = \frc{\mathrm{SL}(3,\fieldR)}{\mathrm{SL}(2,\fieldR) \times
  \mathrm{SO}(1,1)}\ ,
 \end{equation}
which is \emph{not} a quaternion-K\"ahler manifold. This is not in
contradiction with supersymmetry, since only \emph{Minkowskian}
supersymmtry requires the target space to be quaternionic \cite{BW}. A
brief discussion on the geometry of the space \eqref{SL3} is given in
appendix \ref{SL3R}.

In four dimensions, the same target space can be obtained by applying
the \textbf{c}-map \cite{CFG} to pure $N=2$ Euclidean supergravity
\cite{TvN}. This turns the four bosonic degrees of freedom contained in
the metric and graviphoton into the four scalars of the universal
hypermultiplet and gives rise to the two sign flips. Moreover, the
\textbf{c}-map maps Reissner-Nordstrom black hole solutions to
D-instantons in the universal hypermultiplet, as was shown in
\cite{BGLMM}.

We remark that it is the inverted signs in the Euclidean hypermultiplet
action that make instanton solutions in flat space possible. Indeed, the
trace of the Einstein equation sets the bulk Lagrangian to zero, hence
nontrivial field configurations would require a nonvanishing curvature
scalar if the sigma model part of the Lagrangian were positive definite.
The negative signs in \eqref{EUHM-action} allow for cancellations that
are compatible with $R=0$. Note also that since on the hypermultiplet
side the bulk action vanishes for any solution, the instanton action
comes entirely from the boundary term discussed above. As already
stated, the boundary term \eqref{E-BT} is different from the one
proposed in \cite{GS1}. For this reason, we get different results for
the instanton action, and eventually for the instanton corrected
hypermultiplet moduli space.

What is more important from the point of view of instanton calculations,
is that the Euclidean Lagrangian \eqref{EUHM-action} is no longer
positive definite. In a path integral formulation, this makes the finite
action configurations irrelevant, since the action is not bounded from
below. Moreover, perturbative fluctuations around the instanton yield
diverging non-Gaussian integrals, and the semiclassical approximation
would break down. Similar considerations apply to the $N=2$ tensor
multiplet, whose Euclidean action is not positive definite.

On the other hand, the Euclidean double-tensor multiplet Lagrangian is
bounded from below, since the matrix $M_{ab}$ is positive definite. This
leads to a well-defined semiclassical treatment, in which the instantons
dominate the Euclidean path integral. For this reason, it is important
to perform all calculations on the double-tensor multiplet side, and
after having computed the instanton corrections there, we can dualize to
the hypermultiplet formulation.

\section{Instanton solutions} 

\subsection{Asymptotics} 

Before finding the explicit instanton solutions, it will be useful to
discuss the asymptotic behaviour of the fields that can lead to a finite
action. Since the Euclidean action \eqref{E-DTM} consists of three
positive definite terms, each term individually should integrate to a
finite quantity. For simplicity we consider for the moment flat
four-dimensional space. This determines the following behaviour at
infinity:
 \begin{equation} \label{large-r}
  \phi \rightarrow \phi_\infty + \mathcal{O} \Big( \frc{1}{r^2} \Big)\
  ,\quad \chi \rightarrow \chi_\infty + \mathcal{O} \Big( \frc{1}{r^2}
  \Big)\ ,\quad H_{\mu\nu\rho} \propto \frc{1}{r^3}\ .
 \end{equation}
The asymptotic value of $\phi$ is identified with the four- (or five-)
dimensional string coupling constant,
 \begin{equation}
  g_s \equiv \e^{-\phi_\infty/2}\ .
 \end{equation}
The field strengths determine topological charges, defined by
integrating the tensors $H^a$ over spheres at infinity,
 \begin{equation} \label{HQ}
  \int_{S^{D-1}_\infty}\! H^a = Q^{(a)} \ ,\quad a = 1,2\ .
 \end{equation}
In the dual (hypermultiplet) formulation, topological charges become
Noether charges, corresponding to the Peccei-Quinn symmetries which act
as constant shifts in the Lagrange multipliers $\lambda_a$. These
charges descend from the brane charges in ten or eleven dimensions, and,
in the appropriate units, are expected to be quantized.
 
The Euclidean space we shall concentrate on is actually flat space with
a countable number of points, the locations of the instantons,
excised\footnote{A possible contribution to the action from a
Gibbons-Hawking boundary term will then be absent.},
 \begin{equation} \label{space}
  \M = \fieldR^D - \cup_i\, \{\vec{x}_i\}\ ,
 \end{equation}
such that non-trivial cycles with corresponding charges \eqref{HQ}
exist. Stated differently, in the supergravity approximation it will
typically not be possible to find regular solutions at the locations of
the instantons, as we will explicitly see below. The only singularity
which can still lead to a finite action is a logarithmic singularity in
$\phi$ at the origin,
 \begin{equation} \label{small-r}
  \phi \rightarrow c\, \ln r\ ,
 \end{equation}
for some constant $c$. In our examples below, $\chi$ will tend to a
constant $\chi_0$, and the tensors have the same $1/r^3$ behaviour such
that the charges stay the same when the $H^a$ are integrated around an
infinitesimal sphere around the origin.

\subsection{The Bogomol'nyi bound} 

The Euclidean double-tensor multiplet action \eqref{E-DTM} is positive
semi-definite (apart from the Einstein-Hilbert term). In fact, we can
derive a lower bound by writing it as
 \begin{equation}
  \calL_\mathrm{DT}^E = \d^Dx\, \sqrt{g\,} R + \half *\! \big( N\! *\! H +
  O E \big)^t \w \big( N\! *\! H + O E \big) + (-)^D H^t \w N^t O E\ .
 \end{equation}
Here we have defined
 \begin{equation}
  H = \begin{pmatrix} H^1 \\[2pt] H^2 \end{pmatrix}\ ,\quad E =
  \begin{pmatrix} \d\phi \\[2pt] \e^{-\phi/2}\, \d\chi
 \end{pmatrix}\ ,\quad N = \e^{\phi/2} \begin{pmatrix} 0 & \e^{\phi/2}\,
  \\[2pt] 1 & -\chi \end{pmatrix}\ ,
 \end{equation}
such that $N^t N=M$, and $O$ is some orthogonal (scalar) field-dependent
matrix, whose appearance is due to the fact that $N$ and the zweibein
$E$ are determined only modulo local O(2) transformations.

Clearly, the action is bounded from below by
 \begin{equation} \label{bound}
  S^E \geq \int_{\M} \big( \d^Dx\, \sqrt{g\,} R + (-)^D H^t \w N^t
  O E \big)\ ,
 \end{equation}
where the second term is topological, as it is independent of the
spacetime metric. The bound is saturated by field configurations
satisfying the BPS condition
 \begin{equation} \label{BPS}
  * H = - N^{-1} O E\ .
 \end{equation}
A similar Bogomol'nyi equation was derived for an $N=1$, $D=4$ tensor
multiplet (containing one tensor and one scalar) in \cite{R}.
Notice that, if the matrix $O$ is invariant, this equation transforms
covariantly under \eqref{shift-chi} and \eqref{shift-phi}.

Equation \eqref{BPS} is a proper BPS condition only if it implies the
equations of motion, and this will fix the O(2) degeneracy. It is
easily verified that field configurations satisfying \eqref{BPS} have
vanishing energy-momentum tensors, hence they can exist only in
Ricci-flat spaces. We therefore have to amend our BPS condition by the
equation $R_{\mu\nu}(g)=0$.

For the field equations of the tensors, $\d(M\!*\!H)=0$, to hold we must
have
 \begin{equation} \label{closed}
  \d (N^t O E) = 0\ .
 \end{equation}
This condition also guarantees that the topological term in
\eqref{bound} is closed and hence can locally be written as a total
derivative. As a consequence, it does not contribute to the equations of
motion such that also the field equations for the scalars are guaranteed
to be satisfied. The latter follow from requiring that the solution of
\eqref{BPS} correspond to closed forms for $H^a$,
 \begin{equation}
  \d (N^{-1} O *\! E)=0\ .
 \end{equation}

To determine the O(2) matrices that are compatible with \eqref{closed},
we parametrize $O$ by
 \begin{equation}
  O = \begin{pmatrix} 1 & 0 \\[2pt] 0 & \ep \end{pmatrix}
  \begin{pmatrix} c & -s \\[2pt] s & c \end{pmatrix}\ ,
 \end{equation}
where the functions $c(\phi,\chi)$ and $s(\phi,\chi)$ are constrained
by $c^2+s^2=1$, and $\ep=\pm 1$ for the two components of O(2) with
$\det O=\ep$. Equation \eqref{closed} then gives rise to the
differential equations
 \begin{align}
  0 & = \p_\phi c - \e^{\phi/2} \p_\chi s \notag \\*
  0 & = \p_\phi s + \e^{\phi/2} \p_\chi c - \half (2\ep - 1) s\ .
 \end{align}
We derive the general solution in appendix \ref{appO2}. The result is
that there are \emph{three} distinct BPS conditions corresponding to the
O(2) matrices
 \begin{equation}
  O_{1,2} = \pm \begin{pmatrix} 1 & 0 \\[2pt] 0 & \ep \end{pmatrix}\
  ,\quad O_3 = \pm \frc{1}{|\tau'|} \begin{pmatrix} \re \tau' & -\im
  \tau'\, \\[4pt] \im \tau' & \re \tau' \end{pmatrix}\ ,
 \end{equation}
invariant under both \eqref{shift-chi} and \eqref{shift-phi}. Here
$\tau'=\tau-\chi_0$ with $\tau$ as in \eqref{tau}, $\chi_0$ is a real
integration constant, and the plus and minus signs refer to the
instanton and anti-instanton, respectively.

For these three O(2) matrices the 1-form $N^t OE$ is exact,
 \begin{equation} \label{dY}
  N^t O E = \pm\, \d Y\ ,
 \end{equation}
where modulo an additive constant
 \begin{equation}
  Y_{1,2} = \begin{pmatrix} \ep \chi \\[2pt] \e^\phi - \half \ep \chi^2
  \end{pmatrix}\ ,\quad Y_3 = \half \sqrt{4 \e^{\phi} + (\chi -
  \chi_0)^2\,} \begin{pmatrix} 2 \\ - \chi - \chi_0 \end{pmatrix}\ .
 \end{equation}
It follows that the action for BPS configurations is given by a
topological boundary term
 \begin{equation} \label{top-BT}
  S^E\,|_\mathrm{BPS} = (-)^D \int_\M H^t \w N^t O E = \mp \int_{\p\M}
  Y^t H\ .
 \end{equation}
The instanton action is therefore determined by the charges $Q^{(a)}$
and the values of the fields $\chi$ and $\e^{\phi}$ at the boundaries.
 
It is easy to find the corresponding BPS equation in the dual
hypermultiplet formulation. Using \eqref{dual-rel}, \eqref{BPS},
\eqref{dY} and the fact that $M=N^t N$, we find for the Lagrange
multipliers
 \begin{equation}
  \d\lambda = \pm\, \d Y\ ,
 \end{equation}
such that, up to a constant, the solutions for the two extra scalars are
completely determined in terms of $\phi$ and $\chi$.

\subsection{Solutions and instanton action} 

We can solve the BPS condition \eqref{BPS} for the three possible
matrices $O$. For $O_1=\pm 1$, the condition reads
 \begin{equation} \label{O1_BPS}
  * H = \pm \begin{pmatrix} \chi \tbs{\d} \e^{-\phi} \\[2pt] \d
  \e^{-\phi} \end{pmatrix}\ .
 \end{equation}
Applying $\d*$ to the equation and using the Bianchi identities of $H$,
we find that $\e^{-\phi}$ must be harmonic, and from the first component
it then follows that also $\chi$ satisfies the Laplace equation,
 \begin{equation}
  O_1:\quad \d *\! \d\e^{-\phi} = 0\ ,\quad \d *\! \d\chi = 0\ .
 \end{equation}
As mentioned above, scalars satisfying these conditions will also
solve their field equations.

In the following, we consider for simplicity spherically symmetric
configurations (single instantons) in flat space only. The dilaton
equation of motion is then solved by
 \begin{equation} \label{O1_dil}
  \e^{-\phi} = \e^{-\phi_\infty} + \frc{|Q^{(2)}|}{\Omega_D\,
  r^{D-2}}\ .
 \end{equation}
Here $\Omega_D=(D-2)\mathrm{Vol}(S^{D-1})$, and we have chosen the
location of the instanton ($\vec{x}_1$ in \eqref{space}) as the origin.
The integration constant $Q^{(2)}$ appearing in the solution is  equal
to the topological charge associated with $H^2$, as follows from the
second equation of \eqref{O1_BPS}. The `selfdual' instanton (upper sign
in \eqref{O1_BPS}) is taken for negative $Q^{(2)}$, the `anti-selfdual'
instanton for positive $Q^{(2)}$.

Since, up to proportionality factors, there is only a unique spherically
symmetric harmonic function, $\chi$ must be of the form $\chi=\chi_1
\e^{-\phi}+\chi_0$ with $\chi_0$, $\chi_1$ constant. It then follows
from \eqref{O1_BPS} that $\chi_0$ is determined by
 \begin{equation}
  \chi_0 = \frc{Q^{(1)}}{Q^{(2)}}\ ,
 \end{equation}
and this relation is consistent with the shift symmetries
\eqref{shift-chi} and \eqref{shift-phi}, since the charges transform
non-trivially. The instanton action for $O_1$ is given by
 \begin{equation}
  S_1^E = \mp \int_{\p\M} \big[ \chi H^1 + (\e^{\phi} - \half \chi^2)
  H^2 \big]\ ,
 \end{equation}
where the boundary consists of the disjoint union of two spheres, $\p\M
=S^{D-1}_\infty\,\cup\,S^{D-1}_0$, with radii as indicated. The
terms involving $\chi$ will diverge on $S^{D-1}_0$ since $\chi$ is
harmonic, so in order to obtain a finite action we have to take $\chi=
\chi_0$ constant. This was already anticipated from the asymptotic
behaviour of the fields, discussed in the beginning of this section.
The action then reads
 \begin{equation}
  S_1^E = \frac{\big| Q^{(2)} \big|}{g_s^2}\ .
 \end{equation}
This solution was also found in \cite{BB,GS1}, and should correspond to
the fivebrane wrapping the entire Calabi-Yau \cite{BBS}. The instanton
action is positive and hence does not distinguish instantons from
anti-instantons. Imaginary theta-angle-like terms will have to be added
to make this distinction.
\medskip

Turning to $O_2$, we have the BPS condition
 \begin{equation} \label{H-O2}
  * H = \pm\, \d \begin{pmatrix} \e^{-\phi} \chi \\[2pt] \e^{-\phi}
  \end{pmatrix}\ .
 \end{equation}
Again, $\e^{-\phi}$ is harmonic, and the same now applies to $\e^{-\phi}
\chi$. If one imposes rotational symmetry then
 \begin{equation} \label{O2-sol}
  O_2:\quad \d *\! \d\e^{-\phi} = 0\ ,\quad \chi = \chi_1 \e^{\phi} +
  \chi_0\ ,
 \end{equation}
and from \eqref{H-O2}, it follows again that $Q^{(1)}=\chi_0 Q^{(2)}$.
Notice that the field $\chi$ is now completely regular everywhere, and
interpolates between the boundaries according to
 \begin{equation}
  \Delta \chi \equiv \chi_\infty - \chi_0 = \frc{\chi_1}{g_s^2}\ .
 \end{equation}
The complete solution agrees with the asymptotics derived in
\eqref{large-r} and \eqref{small-r}.

For this solution, with the dilaton again given by \eqref{O1_dil},
the instanton action \eqref{top-BT} then becomes
 \begin{equation} \label{S2-inst}
  S_2^E = \big| Q^{(2)} \big|\, \Big( \frc{1}{g_s^2} + \half\,
  (\Delta \chi)^2 \Big)\ .
 \end{equation}
For the particular case of $\Delta \chi=0$, the solution and instanton
action are the same as for the $O_1$ solution. Notice also that both
terms are positive and invariant under the shift symmetries
\eqref{shift-chi} and \eqref{shift-phi}, as guaranteed by the properties
of the original action. For $\Delta\chi\neq 0$, our instanton solution
is new, and this term in the instanton action does not depend on the
string coupling constant $g_s$. The appearance of $\Delta\chi$ in the
instanton action is one of the new results in this paper. Its presence
was somehow anticipated in \cite{BBS}, and here we have computed it
explicitly.
\medskip

We now turn to $O_3$. The BPS equation for this case reads
 \begin{equation}
  * H = \pm \frc{1}{|\tau'|} \begin{pmatrix} -2\, \d\phi + \e^{-\phi}
  (\chi + \chi_0)\, \d\chi + \chi (\chi - \chi_0)\, \d\e^{-\phi}\,
  \\[4pt] (\chi - \chi_0)\, \d\e^{-\phi} + 2 \e^{-\phi} \d\chi
 \end{pmatrix}\ .
 \end{equation}
We have been unable to find the general solution\footnote{The most
general spherically symmetric solution was later found in
\cite{DdVTV}.}. Instead, let us consider two Ans\"atze for which we can
explicitly solve the equations. First, we set $\chi=2\chi_1\e^{\phi/2}
+\chi_0$. Then the equations simplify to
 \begin{equation}
  * H = \pm 2\, \d \e^{-\phi/2} \begin{pmatrix} \sqrt{1 + \chi_1^2\,}\,
  \\[2pt] 0 \end{pmatrix}\ .
 \end{equation}
It follows that now $\e^{-\phi/2}$ is harmonic, with solution
 \begin{equation}
  \e^{-\phi/2} = \e^{-\phi_{\infty}/2} + \frac{\big|Q^{(1)}\big|}
  {2 \sqrt{1 + \chi_1^2\,}\, \Omega_D\, r^{D-2}}\ .
 \end{equation}
The scalar $\chi$ is then completely regular and interpolates between
the boundaries as
 \begin{equation}
  \Delta \chi = \chi_{\infty} - \chi_0 = \frac{2\chi_1}{g_s}\ .
 \end{equation}
Since $H^2=0$ we have $Q^{(2)}=0$, and for the instanton action we find
 \begin{equation} \label{Sinst_O3_1}
  S_3^E = \big| Q^{(1)} \big|\, \sqrt{\frc{4}{g_s^2} + (\Delta \chi)^2}
  = \big| Q^{(1)} \big|\, \big| \tau'_{\infty} \big|\ ,
 \end{equation}
where $\tau'_{\infty}=(\chi_\infty-\chi_0)+2\I\,\e^{\phi_{\infty}/2}$ is
the value of $\tau'$ at infinity. For $\Delta\chi=0$, a similar solution
was also found in \cite{GS1}. Following the discussion in \cite{BBS}, it
should correspond, from a IIA point of view, to the D2-brane wrapping a
three-cycle in the Calabi-Yau, or to the D1+D3+D5-branes wrapping even
cycles in type IIB\@. Notice again consistency with the symmetries
\eqref{shift-chi} and \eqref{shift-phi}. Observe also that for $\Delta
\chi=0$, the solution is inversely proportional to $g_s$, and is for
small $g_s$ dominating over the fivebrane instanton \eqref{S2-inst}.

As a second Ansatz, consider $\chi=2\chi_1\e^\phi+\chi_0$. This differs
from the first Ansatz in the power of $\e^\phi$. The BPS condition turns
into
 \begin{equation}
  * H = \pm 2\, \d \sqrt{\e^{-\phi} + \chi_1^2\,} \begin{pmatrix} 1 -
  \chi_1 \chi_0 \\[2pt] -\chi_1 \end{pmatrix}\ .
 \end{equation}
Accordingly, the square root must be harmonic, and we find for $\chi_1
\neq 0$ (the case $\chi_1=0$ is included in the previous Ansatz),
 \begin{equation}
  \e^{-\phi} = (h - \chi_1) (h + \chi_1)\ ,\quad h = \sqrt{
  \e^{-\phi_\infty} + \chi_1^2\,} + \Big| \frc{Q^{(2)}}{2\chi_1} \Big|\,
  \frc{1}{\Omega_D\, r^{D-2}}\ .
 \end{equation}
The scalar field $\chi$ is regular everywhere and interpolates between
zero and infinity as
 \begin{equation}
  \Delta \chi = \frac{2\chi_1}{g_s^2}\ .
 \end{equation}
The BPS equation further fixes the constant $\chi_1$ to be
 \begin{equation}
  \chi_1 = - \frc{Q^{(2)}}{Q^{(1)} - \chi_0 Q^{(2)}}\ ,
 \end{equation}
and the instanton action is easily computed from \eqref{top-BT},
 \begin{equation}
  S_3^E = \big| \tau'_\infty \big|\, \Big( \big| \hat{Q}^{(1)} \big|
  + \half \big| \Delta \chi\, Q^{(2)} \big| \Big)\ .
 \end{equation}
We have redefined the $Q^{(1)}$ charge according to
 \begin{equation}
  \hat{Q}^{(1)} \equiv Q^{(1)} - \chi_0 Q^{(2)}\ ,
 \end{equation}
such that it is invariant under \eqref{shift-chi}. For $Q^{(2)}=0$, the
instanton action then clearly reduces to \eqref{Sinst_O3_1}.

The obtained results for the instanton action carry over to the
hypermultiplet side, because the dualization procedure does not affect
the real part of the instanton action.

\section{Discussion} 

In this paper, we have carried out the first steps of calculating
instanton corrections to the hypermultiplet moduli space. An important
ingredient was to derive a Bogomol'nyi bound for the double-tensor
multiplet Lagrangian, and to solve the corresponding BPS equation. In a
supersymmetric formulation, adapted to Euclidean space, we expect our
instanton solutions to preserve one half of the supersymmetries. A more
general formulation for the double-tensor multiplet Lagrangian,
including the fermions and supersymmetry transformation rules, is
presently under study. This will be important for finding the fermionic
zero modes and eventually for computing instanton corrections to the
relevant correlation functions that determine the hypermultiplet
quantum-geometry. The exact moduli space must be consistent with the
results derived in our paper. In particular, our supergravity instanton
solutions should match with the results obtained from wrapping branes in
the full ten-dimensional string theory. Stated differently, the
universal hypermultiplet metric must contain exponential corrections
which, at leading order in the string coupling constant and $\alpha'$,
agree with the form of our instanton action. Using some results about
quaternionic geometry \cite{CP,DWRV}, it should be possible to find
quaternionic metrics which asymptotically reproduce our results. We
intend to report further on these issues in the near future.

\section*{Acknowledgements} 

We thank Michael Gutperle and Thomas Mohaupt for discussions and reading
an earlier draft of this paper. U.T.\ thanks the Deutsche
Forschungsgemeinschaft for financial support.

\appendix 

\section{SL(3,R) / SL(2,R) $\times$ SO(1,1)} 
\label{SL3R}

In this appendix, we discuss some geometrical aspects related to the
sigma model corresponding to \eqref{EUHM-action}, with target space
\eqref{SL3}. The easiest way to study this space is by using the fact
that the Minkowskian version of the universal hypermultiplet moduli
space is both K\"ahler and quaternion-K\"ahler. Since K\"ahler geometry
is simpler to analyze, we will study the coset \eqref{SL3} from the
point of view of K\"ahler geometry. Because of the sign flips compared
to the Minkowskian version, the target space will no longer be K\"ahler.
It is therefore not possible to define complex coordinates together with
a K\"ahler potential that determines the metric. As we show below, it is
still possible to define coordinates and a potential from which the
metric can be computed. To see this, we first define the fields
 \begin{equation}
   a = \sigma + \half \chi \varphi\ ,\quad C_\pm = \half (\varphi \pm
  \chi)\ ,
 \end{equation}
in terms of which the sigma model part of the Euclidean Lagrangian
\eqref{EUHM-action} reads
 \begin{equation}
  \calL_\mathrm{UH}^E = \half |\d\phi|^2 - 2 \e^{-\phi} *\! \d C_+ \w
  \d C_- - \half \e^{-2\phi}\, |\d a + C_+ \tbs{\d} C_-|^2\ .
 \end{equation}
If we further pass to coordinates $u^1_\pm,u^2_\pm\in\fieldR$ via the
relations
 \begin{equation}
  S_\pm = \e^{\phi} \mp a - C_+ C_- = \frc{1 \mp u^1_\pm}{1 \pm
  u^1_\pm}\ ,\quad C_\pm = \frc{u^2_\pm}{1 \pm u^1_\pm}\ ,
 \end{equation}
then the Lagrangian can be written as
 \begin{equation}
  \calL^E_\mathrm{UH} = 2 g_{ij} *\! \d u^i_+ \w \d u^j_-
 \end{equation}
with a metric
 \begin{equation} \label{K+-}
  g_{ij} = - \frc{\p^2}{\p u^i_+ \p u^j_-}\, \ln\! \big( 1 + u^1_+
  u^1_- + u^2_+ u^2_- \big)\ .
 \end{equation}
The $u^i_\pm$ are inhomogeneous coordinates of the coset space
\eqref{SL3}, transforming under $M\in\mathrm{SL}(2,\fieldR)$ as $u_+
\rightarrow M u_+$, $u_-\rightarrow (M^{-1})^tu_-$. We have therefore
identified a potential in terms of real coordinates which determines
the metric components. Such spaces are called para-K\"ahler\footnote{A
more general discussion on para-K\"ahler manifolds, in the context of
Euclidean supergravity coupled to vector multiplets, will be given in
\cite{CHM}.}. The metric for the Minkowskian universal hypermultiplet
moduli space \eqref{UHM-QK} is of the same form as in \eqref{K+-}, but
with $u^i_\pm$ treated as complex coordinates, where $u^i_-=-
\bar{u}^i_+$ under complex conjugation.

\section{Determination of O(2) matrices} 
\label{appO2}

We need to solve the differential equations
 \begin{align}
  0 & = \p_\phi c - \e^{\phi/2} \p_\chi s \label{O2_1} \\[2pt]
  0 & = \p_\phi s + \e^{\phi/2} \p_\chi c - \half (2\ep - 1) s\ ,
	\label{O2_2}
 \end{align}
where $c$ and $s$ are subject to the constraint $c^2+s^2=1$. We first
multiply \eqref{O2_1} by $s$ and \eqref{O2_2} by $c$, respectively, and
use $-s\p s=c\p c$ to write the equations as
 \begin{align}
  0 & = s\, \p_\phi c + \e^{\phi/2} c\, \p_\chi c \label{O2_3} \\[2pt]
  0 & = c\, \p_\phi s + \e^{\phi/2} c\, \p_\chi c - \half (2\ep - 1)
	c s\ . \label{O2_4}
 \end{align}
Multiplying the difference of these equations by $c$ gives
 \begin{equation} \label{O2_5}
  0 = c \big[ c\, \tbs{\p_\phi} s - \half (2\ep - 1) c s \big] =
  \p_\phi s - \half (2\ep - 1)\, (1 - s^2) s\ ,
 \end{equation}
which involves only $s$ and can easily be integrated:
 \begin{equation}
  \frc{s^2}{1 - s^2} = \frc{1 - c^2}{c^2} = f^2(\chi)\, \e^{(2\ep - 1)
  \phi}\ .
 \end{equation}
The positive integration constant $f^2$ may depend on $\chi$. These
expressions we plug into the sum of \eqref{O2_3} and \eqref{O2_4},
 \begin{align}
  0 & = \e^{\phi/2} \p_\chi c^2 + \p_{\phi} (cs) - \half (2\ep - 1) cs
	\notag \\[2pt]
  & = - \frc{2 f\, \e^{(3 - 4\ep)\phi/2}}{(f^2 + \e^{(1 - 2\ep)\phi}
	)^2}\ \big[ \p_\chi f \pm \half (2\ep - 1)\, \e^{(\ep - 1)\phi}
	f^2 \big]\ ,
 \end{align}
where the sign ambiguity originates from taking the square root of
$(cs)^2$. The equation is satisfied if the expression in square brackets
vanishes. For $\ep=-1$, this is only possible if $f=0$ since $f$ is
independent of $\phi$. For $\ep=+1$, which corresponds to $O\in
\mathrm{SO}(2)$, we find
 \begin{equation}
  f = 0 \quad\text{or}\quad f = \pm \frc{2}{\chi - \chi_0}\ ,
 \end{equation}
with $\chi_0$ an integration constant. $f=0$ implies $c=\pm 1$ and $s=
0$. For nontrivial $f$ we obtain (with the relative sign fixed by the
original equations \eqref{O2_1} and $\eqref{O2_2}$)
 \begin{equation}
  c = \pm \frac{\chi - \chi_0}{\sqrt{4 \e^\phi + (\chi - \chi_0)^2\,}
  \,}\ ,\quad s = \pm \frc{2\, \e^{\phi/2}}{\sqrt{4 \e^\phi + (\chi -
  \chi_0)^2\,}\,}\ ,
 \end{equation}
or in terms of $\tau'=(\chi-\chi_0)+2\I\,\e^{\phi/2}$,
 \begin{equation}
  c + \I s = \pm \frc{\tau'}{|\tau'|}\ .
 \end{equation}

\raggedright

\end{document}